\documentclass{article}
\usepackage[utf8]{inputenc}
\usepackage{graphicx}
\usepackage{multicol}
\usepackage{authblk}
\usepackage[margin=0.8in]{geometry}
\usepackage{array}
\usepackage{hyperref}
\usepackage{tcolorbox}
\usepackage{markdown}
\usepackage{upquote}
\usepackage{listings}
\usepackage{fancyvrb}
\usepackage{enumitem}

\lstdefinestyle{base}{
  breaklines=true,
  basicstyle=\tiny\ttfamily\color{black},
  moredelim=**[is][\color{red}]{@}{@},
}

\usepackage[sorting = none, backend = bibtex, style=numeric-comp]{biblatex}
\title{Automated Software Vulnerability Static Code Analysis Using Generative Pre-Trained Transformer Models}

\author[1]{Elijah Pelofske\thanks{E-mail: elijah.pelofske@protonmail.com}}
\author[1]{Vincent Urias}
\author[2,1]{Lorie M. Liebrock}
\affil[1]{Sandia National Laboratories}
\affil[2]{New Mexico Cybersecurity Center of Excellence, New Mexico Tech}

\date{\vspace{-6ex}}

\begin{document}

\maketitle

\begin{abstract}
Generative Pre-Trained Transformer models have been shown to be surprisingly effective at a variety of natural language processing tasks -- including generating computer code. However, in general GPT models have been shown to not be incredibly effective at handling specific computational tasks (such as evaluating mathematical functions). 
In this study, we evaluate the effectiveness of open source GPT models, with no fine-tuning, and with context introduced by the langchain and localGPT Large Language Model (LLM) framework, for the task of automatic identification of the presence of vulnerable code syntax (specifically targeting C and C++ source code). This task is evaluated on a selection of $36$ source code examples from the NIST SARD dataset, which are specifically curated to not contain natural English that indicates the presence, or lack thereof, of a particular vulnerability (including the removal of all source code comments). The NIST SARD source code dataset contains identified vulnerable lines of source code that are examples of one out of the $839$ distinct Common Weakness Enumerations (CWE), allowing for exact quantification of the GPT output classification error rate. A total of $5$ GPT models are evaluated, using $10$ different inference temperatures and $100$ repetitions at each setting, resulting in $5,000$ GPT queries per vulnerable source code analyzed. 
Ultimately, we find that the open source GPT models that we evaluated are not suitable for fully automated vulnerability scanning because the false positive and false negative rates are too high to likely be useful in practice. However, we do find that the GPT models perform surprisingly well at automated vulnerability detection for some of the test cases, in particular surpassing random sampling (for some GPT models and inference temperatures), and being able to identify the exact lines of code that are vulnerable albeit at a low success rate. The best performing GPT model result found was Llama-2-70b-chat-hf with inference temperature of $0.1$ applied to NIST SARD test case 149165 (which is an example of a buffer overflow vulnerability), which had a binary classification recall score of $1.0$ and a precision of $1.0$ for correctly and uniquely identifying the vulnerable line of code and the correct CWE number. 
Additionally, the GPT models are able to, with a rate quantifiably better than random sampling, identify the specific line of source that contains the identified CWE for many of the NIST SARD test cases. 
\end{abstract}

%%%%%%%%%%%%%%%%%%%%%%%%%%%%%
%%%%%%%%%%%%%%%%%%%%%%%%%%%%%

\section{Introduction}
\label{section:introduction}

Generative Pre-Trained Transformer (GPT) models, built using the attention architecture \cite{vaswani2023attention}, have been shown to be surprisingly effective at a variety of information processing tasks including computer code generation and natural language text summarization \cite{chen2021evaluating, brown2020language, openai2023gpt4, touvron2023llama}. GPT models are intended to work very well with language, and are typically referred to as a class of Large Language Model (LLM). Notably however, GPT models can, in certain settings, also perform algorithmic tasks such as learning the greatest greatest common divisor algorithm \cite{charton2024learning}. The surprising capability of GPT models to perform well at tasks involving computer code leads to the natural question of whether GPT models could be used for determine security properties of static software samples. In this study, we examine whether current open source GPT models can be used to correctly identify software vulnerabilities and weaknesses in source code correctly.

The evaluation of our study is based on a set of $36$ labeled examples from the highly credible NIST SARD dataset \cite{227831}. Each source code example in this dataset is meticulously mapped to the Common Weakness Enumeration (CWE) framework \cite{christey2013common, 10.1145/1387830.1387835}, providing a solid foundation for our research on software vulnerability detection. The Common Weakness Enumeration (CWE) is a categorization of known types of software weaknesses, hardware weaknesses, and software vulnerabilities.

Static code vulnerability scanners are an important field of study because they speed up the process of identifying and thus patching software and even potentially hardware vulnerabilities and flaws that cause security problems in computer systems. There are a wide variety of static code analysis scanners that have been developed, usually focusing on specific domains or specific languages \cite{1657940, goseva2015capability, 898880, kaur2020comparative, 1624016, 8745996, livshits2005finding, 10.1145/3533767.3534380, 8622456}. A challenge with static vulnerability analysis is that the outputs typically have high false positive rates because the tools usually use string pattern matching to identify known syntax representing an instance of a specific code flaw. The main advantage of static code analysis is that it is much faster and easier to apply than dynamic program analysis.

Ref. \cite{10179324} studies the task of \emph{repairing} CWE examples using GPT models. Ref. \cite{liu2024source} examined source code vulnerability detection using several code language models (namely BERT variants) on a large dataset of software examples.

Our study introduces novel methods compared to existing research, particularly in the use of highly curated datasets that include the exact vulnerable line of source code. The structured GPT prompt we employ with these GPT models enables the automated extraction of vulnerability detection data from the output, facilitating large-scale, repeatable, accuracy analysis. This dataset combined with structured JSON prompting and parsing allows us to fully leverage the capabilities of GPT models for automated vulnerability detection, requiring no manual processing of the GPT output. In total, we generate and process $100 \cdot 10 \cdot 36 \cdot 5 = 180000$ discrete GPT text outputs (each of these being a separate, independent, and complete text inference computations. All of the GPT model computations are performed locally, on a local GPU compute server, using open source GPT models.

\begin{table*}[ht!]
\centering
\begin{tabular}{ |c|c||c|c|c| }
 \hline
 \hline
 Model name & Reference(s) & Context Length & Architecture type & Model Size \\ 
 \hline
 \hline
 \texttt{Llama-2-70b-chat-hf} & \cite{touvron2023llama} & 4096 Tokens & llama & 70B params \\ 
 \hline
 \texttt{zephyr-7b-alpha} &~\cite{rafailov2023direct} & 32768 Tokens & mistral & 7.24B params \\ 
 \hline
 \texttt{zephyr-7b-beta} &~\cite{rafailov2023direct, tunstall2023zephyr} & 32768 Tokens & mistral & 7.24B params \\ 
 \hline
 \texttt{Mistral-7B-Instruct-v0.1} &~\cite{jiang2023mistral} & 32768 Tokens & mistral & 7.24B params \\ 
 \hline
 \texttt{Turdus} &~\cite{udk_dot_ai_turdus} & 32768 Tokens & mistral & 7.24B params \\ 
 \hline
\end{tabular}
\caption{Summary of the $6$ Generative Pre-trained Transformers models used in this study. Here B denotes the order of magnitude billion. }
\label{table:GPT_model_summary}
\end{table*}

%%%%%%%%%%%%%%%%%%%%%%%%%%%%%%%%%%%%%%%%%%%%%%%%%%%%%%%%%%%%%%
%%%%%%%%%%%%%%%%%%%%%%%%%%%%%%%%%%%%%%%%%%%%%%%%%%%%%%%%%%%%%%
%%%%%%%%%%%%%%%%%%%%%%%%%%%%%%%%%%%%%%%%%%%%%%%%%%%%%%%%%%%%%%
\section{Methods}
\label{section:methods}

The five open-source GPT models used in this study are summarized in Table \ref{table:GPT_model_summary}. These GPT models were selected for their general performance among available GPT models - and all of these models can generate and parse computer code. Although a large number of different GPT models could be utilized for this study - due to the high computation time required to perform these experiments, a small number of GPT models were selected. These GPT models are from the huggingface GPT repository \cite{wolf2020huggingfaces} and the GPT models are run using the Python3 modules \texttt{pytorch} \cite{Paszke_PyTorch_An_Imperative_2019} and \texttt{transformers}, using a local compute server that has four Nvidia A100 GPU's~\cite{9361255} (each with 82 Gigabytes of memory), and CUDA Version 12.4.

GPT models can be used to predict the most likely token, given the previous $N$ tokens (where $N$ is the context window for current GPT models on the order of $4096-32768$). When used in this manner, the text produced by the GPT models is deterministic, given that the previous tokens are the same. However, GPT models can learn and generate probability distributions of the most likely subsequent tokens - and we can add stochasticity in the text generation process by sampling from that distribution, weighted by their respective probabilities. The parameter that controls this sampling is known as \emph{temperature}. This stochastic text generation is part of what makes generative machine learning models able to produce a wide variety of outputs. In the context of investigating the usage of these models as code-scanning tools, these models will only perform somewhat well at this task. Moreover, we need outputs to vary somewhat to determine the possible range of characteristics of the output since the text generation process of GPT models can be stochastic. Therefore, we vary the text generation inference temperature from $0.1$ to $1.0$ in increments of $0.1$. The huggingface \cite{wolf2020huggingfaces} python 3 module used to execute this GPT model inference allows temperatures up to $100$, where $0$ denotes deterministic output (with no sampling), and $100$ denotes closer to uniform token sampling. We choose relatively low temperatures because we need the output to conform to parsable data structures. Critically, because the outputs vary significantly, we view this task as a \emph{stochastic sampling} problem, where we characterize the proportion of samples with specific required properties (the most basic being whether the data could be automatically parsed). $100$ samples (GPT outputs) are generated for each inference temperature, GPT model, and source code test case.

\begin{table*}[ht!]
\centering
\begin{tabular}{ |m{1.6cm}||c|m{8.5cm}|c|m{1.9cm}| }
 \hline
 NIST SARD Test Case & CWE id & CWE name & Language & Number of Lines of Source Code \\ 
 \hline
 \hline
 1792 & CWE-79 & Improper Neutralization of Input During Web Page Generation & C & 24 \\ 
 \hline
 500843 & CWE-476 & NULL Pointer Dereference & C++ & 11 \\ 
 \hline
 1779 & CWE-463 & Deletion of Data Structure Sentinel & C & 10 \\ 
 \hline
 1645 & CWE-20 & Improper Input Validation & C & 21 \\ 
 \hline
 149165 & CWE-121 & Stack-based Buffer Overflow & C & 35 \\ 
 \hline
 2015 & CWE-329 & Generation of Predictable IV with CBC Mode & C & 12 \\ 
 \hline
 500757 & CWE-787 & Out-of-bounds Write & C++ & 7 \\ 
 \hline
 149135 & CWE-489 & Active Debug Code & C & 35 \\ 
 \hline
 149203 & CWE-416 & Use After Free & C & 30 \\ 
 \hline
 149111 & CWE-134 & Use of Externally-Controlled Format String & C & 18 \\ 
 \hline
 149185 & CWE-391 & Unchecked Error Condition & C & 11 \\ 
 \hline
 1494 & CWE-74  & Improper Neutralization of Special Elements in Output Used by a Downstream Component & C++ & 12 \\ 
 \hline
 149179 & CWE-401  & Missing Release of Memory after Effective Lifetime & C & 17 \\ 
 \hline
 149107 & CWE-415  & Double Free & C & 55 \\ 
 \hline
 148871 & CWE-188  & Reliance on Data/Memory Layout & C & 77 \\ 
 \hline
 149143 & CWE-120  & Buffer Copy without Checking Size of Input & C & 13 \\ 
 \hline
 149103 & CWE-367  & Time-of-check Time-of-use (TOCTOU) Race Condition & C & 86 \\ 
 \hline
 1501 & CWE-822 & Untrusted Pointer Dereference & C++ & 17 \\ 
 \hline
 149199 & CWE-412  & Unrestricted Externally Accessible Lock & C & 17 \\ 
 \hline
 149183 & CWE-468 & Incorrect Pointer Scaling & C & 10 \\ 
 \hline
 149085 & CWE-244 & Improper Clearing of Heap Memory Before Release & C & 37 \\ 
 \hline
 2079 & CWE-464 & Addition of Data Structure Sentinel & C & 14 \\ 
 \hline
 72 & CWE-248 & Uncaught Exception & C++ & 25 \\ 
 \hline
 2046 & CWE-259 & Use of Hard-coded Password & C++ & 48 \\ 
 \hline
 2060 & CWE-457 & Use of Uninitialized Variable & C++ & 20 \\ 
 \hline
 1989 & CWE-89 & Improper Neutralization of Special Elements used in an SQL Command & C++ & 34 \\ 
 \hline
 114 & CWE-532 & Insertion of Sensitive Information into Log File & C & 16 \\ 
 \hline
 20 & CWE-118 & Incorrect Access of Indexable Resource & C & 10 \\ 
 \hline
 1887 & CWE-22 & Improper Limitation of a Pathname to a Restricted Directory & C & 48 \\ 
 \hline
 313 & CWE-119  & Improper Restriction of Operations within the Bounds of a Memory Buffer & C & 12 \\ 
 \hline
 2016 & CWE-378  & Creation of Temporary File With Insecure Permissions & C & 12 \\ 
 \hline
 149125 & CWE-122 & Heap-based Buffer Overflow & C & 53 \\ 
 \hline
 501035 & CWE-563  & Assignment to Variable without Use & C++ & 11 \\ 
 \hline
 1852 & CWE-170  & Improper Null Termination & C & 13 \\ 
 \hline
 149163 & CWE-99  & Improper Control of Resource Identifiers & C & 23 \\ 
 \hline
 149241 & CWE-78 & Improper Neutralization of Special Elements used in an OS Command & C & 41 \\ 
 \hline
\end{tabular}
\caption{NIST SARD test cases that are used in this study. Each of these target test functions has exactly one line of source code that has been identified as being an example of a CWE. The number of lines of source code field is computed by the absolute number of lines of code syntax from the original dataset, not including lines containing only whitespace, and with all comments having been removed, and with no manual formatting (for example, curly brace placement). }
\label{table:NIST_SARD_Test_cases}
\end{table*}

To ground the GPT model output in the definitions of the Common Weakness Enumeration (CWE) framework, we utilize a technique known as retrieval augmented generation \cite{lewis2021retrievalaugmented}. Specifically, we use a codebase known as \texttt{localGPT} \cite{localGPT} that uses langchain \cite{Chase_LangChain_2022} to augment the capabilities of GPT models by using additional word embeddings of reference documents (not as part of the model pre-training). Specifically, the source documents are split into character maximum chunk sizes of 1000, with chunk overlap of 200 characters. 

The localGPT prompt template is a combination of context and then the user prompt question. No chat history was used, meaning that each text generation call is independent. The prompt construction uses a system prompt \cite{localGPT} followed subsequently by the context of the generated word embeddings and then the user prompt. The system prompt used in this study, which is from a version of the localGPT codebase \cite{localGPT}, is given below:

\begin{tcolorbox}
You are a helpful assistant, you will use the provided context to answer user questions.
Read the given context before answering questions and think step by step. If you can not answer a user question based on the provided context, inform the user. Do not use any other information for answering user. Provide a detailed answer to the question.
\end{tcolorbox}

The reference documents that are used to provide context for the GPT model inference calls are the following: one text file for each of the $839$ CWEs that are currently defined, where the text file contains the CWE number, the name of the software weakness, and the complete description of the CWE from Mitre \footnote{\url{https://cwe.mitre.org}}. Next is a text file with a brief definition of the CWE framework. And lastly, two books (PDF files of the books) on software, security, and vulnerabilities (specifically for the C family languages) \cite{kernighan2002c, dowd2006art}.

Table \ref{table:NIST_SARD_Test_cases} outlines the exact code test cases that are analyzed in this study. Note that NIST SARD test case 148871 is the vulnerable code that was tracked by CVE-2009-3549, and test case 149199 is an example of an implementation of CVE-2002-1869. All of the selected NIST SARD test cases have a few specific properties that we manually selected these for. First, each test case is an example of a different CWE, which makes the results representative of a variety of different possible software weaknesses. Second, each test case contains exactly one line of source code that is an example of a particular CWE; this simplifies the parsing and accuracy metrics. Third, each test case contains no variable names, text, or function names that indicate whether the source code contains a vulnerability or not. In cases where there is a string or function name that indicates what type of example the source code is, we manually remove, or change, that string before parsing by the GPT models. Fourth, for no test cases does the vulnerable line of code occur more than once in the entire source code; this eliminates ambiguity for measuring the accuracy rate of the GPT models for identifying specific lines of source code. Any comments in the source code examples are removed before parsing by the GPT models so as to not give the GPT models natural language that describes the test cases. 

Appendix \ref{section:appendix_NIST_SARD_test_cases} shows the exact source code for $10$ of these NIST SARD test cases. The phrase \emph{vulnerable line of source} will be used throughout the text to refer to the exact syntax of the vulnerable line of code -- not line number. Line numbers can be somewhat ambiguous, especially for these C syntax test cases, and moreover having the GPT model output the exact source code makes checking the validity of that output much clearer.

The exact prompt that is used for all of the GPT models is given in the text box below. Note the use of the triple backticks to delineate where the source code is, and where the JSON datastructure is. The use of the triple backticks in the prompt allows the GPT outputs to be automatically parsed by using the triple backticks, and possibly the markdown code identifier, as the delimiter. Note that this triple backtick parsing does not always work, in particular if the JSON is not correctly formatted or the triple backticks are not correctly used in the GPT output. The prompt is intended to be very source code agnostic; there are no details given of the source code in the prompt. The prompt is also intended to not bias the output towards specifying a specific number of vulnerabilities. In the case of the $36$ NIST SARD test cases used in this study, the number of CWEs is always exactly $1$, however the goal is for this method to more general than only identifying a single CWE and moreover we want the GPT model to be able to fail in a variety of ways in order to rigorously evaluate its accuracy (in this case, identifying more CWEs than are present would constitute a high false positive rate).

\begin{tcolorbox}
I am seeking assistance to analyze provided source code for the purpose of detecting potential Common Weakness Enumeration (CWE) vulnerabilities. 
The key requirement is for the analysis results to be exclusively formatted in JSON enclosed by triple backticks, listing only the CWEs that are detected as 'Present' in the code, thereby omitting any 'Not Present' entries for a cleaner output. 
The exact line, or lines, of code which have this CWE vulnerability needs to be specified in the returned JSON structure, along with a succinct description of what the vulnerability category is for this source code. 
The JSON output should enumerate each identified CWE with its identification number, the exact source code syntax that causes the CWE, and a description of the vulnerability, as shown in the example structure below:

\begin{verbatim}
```json
{"findings": [
	{
		"CWE_Number": "CWE-XX",
		"Status": "Present",
		"Source_Code": "Vulnerable source code syntax",
		"Description:" "Summary of the category of software vulnerability, 
        and why this code is vulnerable"
	},
	{
		"CWE_Number": "CWE-YY", 
		"Status": "Present",
		"Source_Code": "Vulnerable source code syntax",
		"Description:" "Summary of the category of software vulnerability, 
        and why this code is vulnerable"
	}
]
}
```
\end{verbatim}

Source code:
\begin{verbatim}
```
--Source Code--
```
\end{verbatim}

Re-check that your output contains the required JSON data-structure before emitting text. 
\end{tcolorbox}

The actual GPT prompt is generated by replacing the vulnerable source code in place of the text \texttt{--Source Code--}.

The usage of requesting the vulnerable code syntax and the description of the vulnerability has two uses:

\begin{enumerate}[noitemsep]
    \item Ground the GPT response to only respond based on the provided source code
    \item To check whether the provided syntax is correct. Because of potential ambiguity with line numbers, only exact syntax string matches are checked for, not integer line number responses. 
\end{enumerate}

The \emph{description} field is not parsed for accuracy or correctness, but the text contained in that field is an interesting indicator of how correct (or incorrect) the GPT output is. For this reason, when reporting results (in Section \ref{section:results}), we provide specific instances of GPT outputs, including the description fields.

Some of the raw GPT output includes characters that are not ascii and can cause issues with saving and parsing the output. Therefore, the raw GPT output is encoded as utf-8 where character parsing errors are ignored and decoded into a parsable string using utf-8. The field in the JSON output \texttt{Source\_Code} has repeated excess whitespace removed, as is done with the reference vulnerable line of source code from each of the NIST SARD samples; this ensures that when the string equality check is performed, the only relevant characteristic with respect to whitespace in the string that is being checked is that the whitespace is delineating characters (for example, repeated spaces, tabs, newlines do not impact this accuracy check). 

The GPT output is (attempted to be) parsed in several stages, as follows:

\begin{enumerate}[noitemsep]
    \item First, if there are triple backtick delimiters in the output, then the output is split at those characters and the first enclosed set of strings is attempted to be parsed as a valid JSON structure. Interestingly, occasionally the GPT model will add in markdown code identifiers immediately after the triple backticks - in these cases we automatically check for these and remove them if they occur so that the structure can be parsed. These markdown identifiers are given in Appendix \ref{section:appendix_markdown_code_identifiers}. 
    \item If the previous step failed, then we attempt an unstructured JSON parsing approach, where output strings are incrementally deleted from the beginning and from the end until a JSON structure can be parsed (or, not). 
\end{enumerate}

Using this parsing, any valid JSON strings that are produced as part of the GPT output will be automatically parsed. This allows this study to perform a systematic evaluation of the accuracy rates of the GPT inference output. This type of data parsing is also highly scalable to a large number of static code samples - and importantly does not require any manual input, tuning, or processing. In other words, this analysis pipeline is entirely \emph{automated}.

\subsection{CWE Identification Accuracy Measures}
\label{section:methods_accuracy_measures}

This section lists several measures for quantifying the success rate of the GPT models for identifying known vulnerable source code.

The first four metrics that are checked are defined as follows:

\begin{enumerate}[noitemsep]
    \item Count of parsable GPT outputs (being able to extract a JSON structure and parse it into Python, with not necessarily correct JSON fields. For example a JSON with an empty list is a valid parse. However, this step does require that there is at least one JSON key, which is correctly named as \texttt{findings}. The subsequent value is allowed to be empty or have any sort of incorrect structure (which could denote that the model is returning no found CWEs) but we need at least this key to validate that a JSON datastructure was parsed. 
    \item The CWE field is labeled as \texttt{Present} is correct (meaning, the correct CW number is listed), but the identified source code is not correct (meaning, it is not the labeled line of code that has the vulnerability) and does not contain the vulnerable line of code either. 
    \item The CWE is correct and identified as \texttt{Present} and the vulnerable code is contained within the identified code syntax but also includes other lines of code from the test case that are not themselves vulnerable as defined by the NIST SARD dataset. These instances can be interpreted as being partially correct. 
    \item Count of fully correct instances where the CWE is correct (and identified as \texttt{Present}) and the identified line of code is correctly identified as the vulnerable code. 
\end{enumerate}

This last metric (metric number 4) is the most important accuracy measure because it counts out of the $1,000$ GPT outputs how many correctly performed the computation that we are attempting to prompt the GPT model to perform. 

For none of these four metrics is the Description field, or any other additional field in the JSON, parsed for correctness or even considered. All of these measures are integers, where the larger the integer is the better the GPT model performed, but the absolute correct metric is measure 4. Note that the quantities measured in metrics 2, 3, and 4 (in the immediately preceding list of measures) are always disjoint; there are no instances that are counted in two or more of these measures. An important, and informative, aspect of these metrics is finding cases where the GPT output correctly identified the vulnerable line of code. However, since line numbers of the code are not checked or requested in the prompt (since they are somewhat ambiguously defined especially for C-family languages), a potential source of error in this quantification is if there exist multiple lines of code with the same syntax, but only one is vulnerable. This does happen, but only in one of the test cases. This is NIST SARD test case 149107, which is CWE-415 (Double Free) which by its definition contains two lines of memory free calls. Therefore, for this one test instance the identification of the vulnerable line of code is ambiguous, but for all others the unique identification is valid and correct.

Another set of metrics that are measured are binary classification error rates. This is computed by considering the GPT output for identified CWEs as a vector of True or False entries, with length equal to the number of valid CWEs (which is currently $839$). This measure disregards both the requested \texttt{Source\_Code} and \texttt{Description} fields (and in particular does not consider whether these entries even exist in the JSON structure. For all \texttt{CWE\_Number} entries (that identifies a valid CWE number), and whose \texttt{Status} is exactly \texttt{Present} (allowing for different capitalizations), that is measured as a binary state of True. All other entries (including instances of identified CWE numbers that are not valid) are measured as binary classifications of False. JSON dictionaries that are completely empty are a valid GPT output that is identifying no CWEs (meaning a binary classification vector of length $839$ that is all False entries). These measures are only computed on the outputs that can be parsed; all outputs that can not be parsed are not counted towards any classification accuracy.\footnote{In the case where a JSON entry contains multiple entries with the same CWE number, that case is treated as the GPT model having identified that CWE number only once; in other words, repeated entries are ignored. This specifically works for this accuracy metric of binary classification since all other fields (except \texttt{Status} are ignored. } These binary classification measures allow rates of false negatives, false positives, true negatives, and true positives to be measured, and then standard machine learning classification measures such as \emph{recall} and \emph{precision} can thus be computed. Precision is intuitively defined as the classifier (which in this case is a GPT language model) being able to not label as positive a sample that is negative. Recall is intuitively defined as the classifier (which in this case is a GPT language model) being able to find all of the positive samples. These metrics being $0$ corresponds to the highest possible error rate, and $1$ is the best measure these metrics can be. 

All parsing steps require the field names in the JSON datastructure to be correct (correct spelling, no ancillary text), but differing capitalizations are always allowed.

%%%%%%%%%%%%%%%%%%%%%%%%%%%%%%%%%%%%%%%%%%%%%%%%%%%%%%%%%%%%%%
%%%%%%%%%%%%%%%%%%%%%%%%%%%%%%%%%%%%%%%%%%%%%%%%%%%%%%%%%%%%%%
%%%%%%%%%%%%%%%%%%%%%%%%%%%%%%%%%%%%%%%%%%%%%%%%%%%%%%%%%%%%%%
\section{Results}
\label{section:results}

For each of the $36$ NIST SARD test cases in Table \ref{table:NIST_SARD_Test_cases}, a total of $5,000$ inference calls are performed using the methods described in Section \ref{section:methods} for the $5$ different open-sourced GPT models. The accuracy measures from these outputs are then summarized using several high-level metrics. Tables \ref{table:accuracy_measures_id_source_code} and \ref{table:recall_and_precision_measures} present two different accuracy measure summaries. Table \ref{table:accuracy_measures_id_source_code} counts how many of the GPT outputs were parsable and correctly identified the CWE number and vulnerable line of code. Table \ref{table:recall_and_precision_measures} shows the binary classification measures recall and precision. Table \ref{table:code_line_identification_only} reports the absolute integer count of how many of the GPT outputs correctly identified the single line of vulnerable source code; importantly, these counts are independent of correct identification of the relevant CWE. These counts, as a proportion out of $1000$ should be compared against uniform random sampling of the total lines of source code (shown in Table \ref{table:NIST_SARD_Test_cases}). Table \ref{table:code_line_identification_only} shows that the correct identification of vulnerable lines of code is much higher than the correct identification of the CWE number and the vulnerable line of source code, as measured in Table \ref{table:accuracy_measures_id_source_code}. Notably in Table \ref{table:code_line_identification_only}, only for a few of the NIST SARD test cases is the count of ever correctly identifying the syntax of vulnerable line of source code zero across all GPT models and settings; namely test cases 149203, 148871, and 2046. For some of the test cases, the proportion of correctly identified vulnerable line of source code is very high. The test cases where the vulnerable line of source was not uniquely identified in Table \ref{table:code_line_identification_only} does not necessarily mean that the correct CWE identification was always 0 (when analyzing strictly whether the identified CWE was correct) - an example of this is test case 2046. 

Table \ref{table:recall_and_precision_measures} shows that for the code test cases where the accuracy is $0$, the $5$ different language models do not perform uniformly bad -- for some of the models the accuracy is always 0 for a given test case, but then when analyzed with a different model the accuracy is non-zero (an example of this is NIST SARD test case 500757). This means that some of the GPT models do perform better than other GPT models for specific test cases. And moreover, there is not consistently a single model that always outperforms the other models.

Table \ref{table:best_performing_recall_and_precision_measures} shows the best performing GPT model and inference temperature in terms of recall and precision measures. Finally, Table \ref{table:most_common_incorrect_CWE_ids} counts the most frequently identified false positives. The best performing GPT model and inference temperature shows that there is no specific model, or temperature setting, that consistently performs the best - and in in particular, every one of the tested GPT models was the best performer for at least one of the CWE test cases. This shows that this task does require diversity of the model architecture, and very likely the type of text on which the models were trained. Lastly, even with these best performing GPT models and inference temperatures some of the CWE test cases were never correctly identified, meaning that the accuracy rates for all GPT models and settings were always 0 - that happened for $11$ out of $36$ of the test cases as denoted by asterisks in Table \ref{table:best_performing_recall_and_precision_measures}.

Section \ref{section:results_example_correct_JSON} then gives several examples of correct JSON output from the GPT models where the vulnerable line of source code was identified, and where the description is correct in some cases and incorrect in other cases.

Table \ref{table:most_common_incorrect_CWE_ids} quantifies the exact counts of false positive CWE id occurrences for the full matrix of the $5$ GPT models and NIST SARD test cases. Specifically, for each NIST SARD test case and GPT model, Table \ref{table:most_common_incorrect_CWE_ids} reports the most frequently mis-identified CWE number. Table \ref{table:most_common_incorrect_CWE_ids} demonstrates two important trends. First, the proportion of false positives is \emph{very} high, and this is consistent across the different GPT models. Second, for most test cases and GPT models the distribution of incorrect CWE ids is not uniformly distributed across the possible CWE numbers - it is heavily biased towards a few specific CWE numbers. For example, \texttt{Mistral-7B-Instruct-v0.1} shows a clear bias towards predicting CWE-124 for multiple test cases. Similarly, \texttt{Llama-2-70b-chat-hf} shows a consistent bias for (incorrectly) predicting CWE-123, and \texttt{Turdus} shows a consistent bias for (incorrectly) predicting CWE-476.

\begin{table*}[h!]
\centering
\scalebox{0.95}{
\begin{tabular}{ |m{3.0cm}||c|c|c|c|c| }
 \hline
 NIST SARD Test Case & \scriptsize{Llama-2-70b-chat-hf} & \scriptsize{zephyr-7b-alpha} & \scriptsize{zephyr-7b-beta} & \scriptsize{Mistral-7B-Instruct-v0.1} & \scriptsize{Turdus} \\ 
 \hline
 1792 (CWE-79) & 937, 1, 0, 0 & 82, 0, 0, 0 & 66, 0, 0, 1 & 430, 1, 0, 0 & 540, 0, 0, 3 \\ 
 \hline
 500843 (CWE-476) & 953, 9, 0, 2 & 66, 4, 0, 1 & 164, 1, 0, 1 & 378, 0, 0, 0 & 554, 24, 59, 14 \\ 
 \hline
 1779 (CWE-463) & 801, 0, 0, 0 & 62, 0, 0, 0 & 85, 0, 0, 0 & 280, 0, 0, 0 & 413, 0, 0, 0 \\ 
 \hline
 1645 (CWE-20) & 668, 7, 0, 0 & 278, 4, 0, 1 & 147, 15, 4, 4 & 704, 0, 0, 0 & 738, 8, 0, 1 \\ 
 \hline
 149165 (CWE-121) & 927, 3, 0, 325 & 183, 9, 1, 4 & 155, 1, 4, 10 & 734, 0, 0, 1 & 597, 0, 0, 0 \\ 
 \hline
 2015 (CWE-329) & 973, 0, 0, 0 & 118, 0, 0, 0 & 116, 0, 0, 0 & 345, 0, 0, 0 & 462, 0, 0, 0 \\ 
 \hline
 500757 (CWE-787) &  993, 0, 0, 0 & 128, 1, 0, 0 & 324, 0, 0, 0 & 308, 0, 0, 0 & 564, 1, 35, 12 \\ 
 \hline
 149135 (CWE-489) & 800, 0, 0, 0  & 82, 0, 0, 0 & 74, 0, 0, 0 & 575, 0, 0, 0 & 514, 0, 0, 0 \\ 
 \hline
 149203 (CWE-416) & 757, 0, 0, 0  & 174, 1, 0, 0 & 222, 0, 0, 0 & 572, 0, 0, 0 & 532, 4, 0, 0 \\ 
 \hline
 149111 (CWE-134) & 718, 297, 3, 56  & 286, 1, 0, 0 & 143, 0, 0, 0 & 672, 7, 0, 2 & 531, 0, 0, 1 \\ 
 \hline
 149185 (CWE-391) &  774, 0, 0, 0  & 286, 0, 0, 0 & 197, 0, 0, 0 & 685, 0, 0, 0 & 531, 0, 0, 0 \\ 
 \hline
 1494 (CWE-74) &  986, 0, 0, 0 & 208, 0, 0, 0 & 171, 0, 0, 0 & 689, 1, 0, 1 & 818, 0, 1, 0 \\ 
 \hline
 149179 (CWE-401) &  981, 5, 0, 0 & 207, 1, 0, 0 & 152, 4, 0, 2 & 599, 0, 0, 0 & 501, 2, 0, 0 \\ 
 \hline
 149107 (CWE-415) &  556, 0, 0, 0 & 228, 0, 0, 0 & 65, 0, 0, 0 & 664, 0, 0, 0 & 431, 0, 0, 0 \\ 
 \hline
 148871 (CWE-188) & 693, 0, 0, 0  & 153, 0, 0, 0 & 30, 0, 0, 0 & 474, 0, 0, 0 & 389, 0, 0, 0 \\ 
 \hline
 149143 (CWE-120) & 987, 14, 1, 27  & 213, 3, 4, 7 & 157, 0, 0, 0 & 740, 2, 0, 0 & 587, 4, 4, 52 \\ 
 \hline
 149103 (CWE-367) & 810, 0, 0, 0  & 71, 0, 0, 0 & 11, 0, 0, 0 & 387, 0, 0, 0 & 647, 0, 0, 0 \\ 
 \hline
 1501 (CWE-822) &  714, 0, 0, 0 & 168, 0, 0, 0 & 105, 0, 0, 0 & 695, 0, 0, 0 & 499, 0, 0, 0 \\ 
 \hline
 149199 (CWE-412) & 945, 0, 0, 0  & 285, 0, 0, 0 & 191, 0, 0, 0 & 701, 0, 0, 0 & 435, 1, 0, 0 \\ 
 \hline
 149183 (CWE-468) & 997, 0, 0, 0  & 275, 0, 0, 0 & 280, 0, 0, 0 & 737, 0, 0, 0 & 638, 0, 0, 0 \\ 
 \hline
 149085 (CWE-244) & 918, 0, 0, 0  & 89, 0, 0, 0 & 151, 0, 0, 0 & 539, 0, 0, 0 & 677, 0, 0, 0 \\ 
 \hline
 2079 (CWE-464) & 984, 0, 0, 0  & 245, 0, 0, 0 & 213, 0, 0, 0 & 568, 0, 0, 0 & 628, 0, 1, 0 \\ 
 \hline
 72 (CWE-248) & 877, 0, 0, 0  & 47, 0, 0, 0 & 193, 0, 0, 0 & 286, 0, 0, 0 & 528, 0, 0, 0 \\ 
 \hline
 2046 (CWE-259) & 885, 0, 0, 0  & 168, 1, 0, 0 & 146, 0, 0, 0 & 561, 0, 0, 0 & 700, 0, 0, 0 \\ 
 \hline
 2060 (CWE-457) &  545, 0, 0, 0 & 70, 0, 0, 0 & 110, 0, 0, 0 & 359, 0, 0, 0 & 559, 0, 0, 0 \\ 
 \hline
 1989 (CWE-89) &  837, 56, 2, 358 & 80, 30, 0, 12 & 78, 11, 6, 5 & 361, 0, 0, 0 & 357, 125, 10, 3 \\ 
 \hline
 114 (CWE-532) &  987, 0, 0, 0 & 318, 2, 0, 0 & 186, 0, 0, 0 & 767, 0, 0, 0 & 625, 0, 0, 0 \\ 
 \hline
 20 (CWE-118) & 410, 0, 0, 0  & 158, 0, 0, 0 & 139, 0, 0, 0 & 424, 0, 0, 0 & 391, 0, 0, 0 \\ 
 \hline
 1887 (CWE-22) & 948, 1, 0, 0  & 103, 0, 0, 0 & 62, 0, 0, 0 & 528, 0, 0, 0 & 577, 1, 0, 0 \\ 
 \hline
 313 (CWE-119) &  991, 0, 0, 1 & 308,i 2, 0, 1 & 138, 0, 0, 0 & 616, 72, 0, 4 & 739, 0, 0, 3 \\ 
 \hline
 2016 (CWE-378) & 428, 0, 0, 0  & 95, 0, 0, 0 & 178, 0, 0, 0 & 387, 0, 0, 0 & 487, 0, 0, 0 \\ 
 \hline
 149125 (CWE-122) & 852, 0, 0, 0  & 220, 1, 0, 7 & 87, 0, 0, 0 & 667, 0, 0, 0 & 477, 0, 1, 3 \\ 
 \hline
 501035 (CWE-563) & 851, 0, 0, 0  & 238, 0, 0, 0 & 123, 0, 0, 0 & 502, 0, 0, 0 & 290, 0, 0, 0 \\ 
 \hline
 1852 (CWE-170) & 986, 0, 0, 0  & 236, 0, 0, 0 & 202, 0, 0, 0 & 683, 0, 0, 0 & 670, 0, 0, 0 \\ 
 \hline
 149163 (CWE-99) & 803, 0, 0, 0  & 227, 0, 0, 0 & 95, 0, 0, 0 & 609, 0, 0, 0 & 574, 0, 0, 0 \\ 
 \hline
 149241 (CWE-78) & 956, 49, 0, 0  & 100, 7, 0, 2 & 96, 5, 0, 0 & 149, 0, 0, 0 & 323, 15, 1, 4 \\ 
 \hline
\end{tabular}}
\caption{Summary of the four accuracy metrics that take into account the identification, or failure to identify, the vulnerable source code, for each of the $5$ GPT models. The four metrics are given as list of integers in each entry in the table. These measures, for each entry in the table, are taken from the $10$ different inference temperature settings used for each model, meaning all four quantities reported in each cell are out of distinct $1,000$ GPT outputs. The accuracy metrics are described in Section \ref{section:methods_accuracy_measures}, but briefly the first integer is a count of how many of the GPT outputs could be parsed into a datastructure, the second integer is that the identified CWE number was correct (and uniquely identified) but the identified line of code was not correct, the third integer is that the CWE number was correct (and uniquely identified) but the identified lines of code included the vulnerable line of code plus additional syntax from the source code, and the fourth integer is a count of how many of the GPT outputs correctly and uniquely identified both the CWE number and the vulnerable line of code. }
\label{table:accuracy_measures_id_source_code}
\end{table*}

\begin{table*}[h!]
\centering
\scalebox{1.0}{
\begin{tabular}{ |m{3.0cm}||c|c|c|c|c| }
 \hline
 NIST SARD Test Case & \scriptsize{Llama-2-70b-chat-hf} & \scriptsize{zephyr-7b-alpha} & \scriptsize{zephyr-7b-beta} & \scriptsize{Mistral-7B-Instruct-v0.1} & \scriptsize{Turdus} \\ 
 \hline
 \hline
 1792 (CWE-79) & 0.022, 0.019  & 0, 0 & 0.015, 0.01 & 0.005, 0.01 & 0.026, 0.021 \\ 
 \hline
 500843 (CWE-476) & 0.012, 0.011  & 0.119, 0.101 & 0.055, 0.034 & 0, 0 & 0.224, 0.393 \\ 
 \hline
 1779 (CWE-463) & 0, 0 & 0, 0 & 0, 0 & 0, 0 & 0, 0 \\ 
 \hline
 1645 (CWE-20) & 0.225, 0.159  & 0.062, 0.047 & 0.435, 0.299 & 0, 0 & 0.008, 0.008 \\ 
 \hline
 149165 (CWE-121) &  0.386, 0.283 & 0.084, 0.072 & 0.221, 0.16 & 0.001, 0.002 & 0, 0 \\ 
 \hline
 2015 (CWE-329) &  0, 0 & 0, 0 & 0.008, 0.006 & 0, 0 & 0, 0 \\ 
 \hline
 500757 (CWE-787) & 0, 0  & 0.023, 0.022 & 0.012, 0.009 & 0, 0 & 0.091, 0.096 \\ 
 \hline
 149135 (CWE-489) &  0, 0 & 0, 0 & 0, 0 & 0, 0 & 0, 0 \\ 
 \hline
 149203 (CWE-416) &  0.0520, 0.016 & 0.139, 0.07 & 0.108, 0.046 & 0.006, 0.007 &  0.017, 0.012 \\ 
 \hline
 149111 (CWE-134) &  0.545, 0.454 & 0.011, 0.01 & 0.007, 0.006 & 0.041, 0.069 & 0.002, 0.002 \\ 
 \hline
 149185 (CWE-391) & 0, 0  & 0, 0 & 0, 0 & 0, 0 & 0, 0 \\ 
 \hline
 1494 (CWE-74) & 0, 0  & 0, 0 & 0, 0 & 0.003, 0.005 & 0.001, 0.001 \\ 
 \hline
 149179 (CWE-401) & 0.005, 0.005  & 0.024, 0.022 & 0.171, 0.117 & 0, 0 & 0.011, 0.011 \\ 
 \hline
 149107 (CWE-415) & 0, 0  & 0.018, 0.007 & 0, 0 & 0.002, 0.001 & 0, 0 \\ 
 \hline
 148871 (CWE-188) & 0, 0  & 0, 0 & 0, 0 & 0, 0 & 0, 0 \\ 
 \hline
 149143 (CWE-120) & 0.048, 0.041  & 0.094, 0.094 & 0, 0  & 0.009, 0.012 & 0.123, 0.125 \\ 
 \hline
 149103 (CWE-367) & 0.001, 0  & 0.014, 0.01 & 0, 0 & 0, 0 & 0, 0 \\ 
 \hline
 1501 (CWE-822) & 0, 0  & 0, 0 & 0, 0 & 0.001, 0.002 & 0, 0 \\ 
 \hline
 149199 (CWE-412) & 0, 0  & 0, 0 & 0, 0 & 0.003, 0.003 & 0.002, 0.002 \\ 
 \hline
 149183 (CWE-468) &  0, 0 & 0, 0 & 0, 0 & 0, 0 & 0, 0 \\ 
 \hline
 149085 (CWE-244) & 0, 0  & 0, 0 & 0, 0 & 0, 0 & 0, 0 \\ 
 \hline
 2079 (CWE-464) &  0, 0 & 0, 0 & 0, 0 & 0, 0 & 0.003, 0.003 \\ 
 \hline
 72 (CWE-248) & 0, 0  & 0, 0 & 0, 0 & 0, 0 & 0, 0 \\ 
 \hline
 2046 (CWE-259) &  0.148, 0.039 & 0.024, 0.016 & 0.027, 0.015 & 0.005, 0.004 & 0.001, 0.001 \\ 
 \hline
 2060 (CWE-457) &  0.002, 0.001 & 0, 0 & 0.009, 0.006 & 0, 0 & 0.002, 0.002 \\ 
 \hline
 1989 (CWE-89) & 0.602, 0.586  & 0.725, 0.617 & 0.654, 0.415 & 0.008, 0.011 & 0.762, 0.665 \\ 
 \hline
 114 (CWE-532) & 0, 0  & 0.006, 0.006 & 0, 0 & 0, 0 & 0, 0 \\ 
 \hline
 20 (CWE-118) & 0, 0  & 0, 0 & 0, 0 & 0, 0 & 0, 0 \\ 
 \hline
 1887 (CWE-22) &  0.657, 0.186 & 0.01, 0.006 & 0, 0 & 0, 0 & 0.003, 0.002 \\ 
 \hline
 313 (CWE-119) & 0.001, 0.001  & 0.019, 0.019 & 0, 0 & 0.136, 0.16 & 0.015, 0.015 \\ 
 \hline
 2016 (CWE-378) & 0, 0  & 0, 0 & 0, 0 & 0, 0 & 0, 0 \\ 
 \hline
 149125 (CWE-122) &  0.015, 0.005 & 0.086, 0.053 & 0.149, 0.082 & 0.004, 0.005 & 0.038, 0.028 \\ 
 \hline
 501035 (CWE-563) &  0, 0 & 0, 0 & 0, 0 & 0, 0 & 0, 0 \\ 
 \hline
 1852 (CWE-170) & 0, 0  & 0, 0 & 0, 0 & 0, 0 & 0, 0 \\ 
 \hline
 149163 (CWE-99) & 0, 0  & 0, 0 & 0, 0 & 0.002, 0.002 & 0, 0 \\ 
 \hline
 149241 (CWE-78) &  0.801, 0.439 & 0.12, 0.074 & 0.073, 0.04 & 0, 0 & 0.087, 0.104 \\ 
 \hline
\end{tabular}}
\caption{Recall and Precision binary classification metrics for each of the $5$ GPT models, taken as the complete vector of $100$ inference calls for each of the $10$ temperature settings. Of these $1,000$ GPT outputs, only those that were able to be parsed contribute to the recall and precision accuracy measures. All quantities are rounded to 3 decimal places. }
\label{table:recall_and_precision_measures}
\end{table*}

\begin{table*}[h!]
\centering
\scalebox{1.0}{
\begin{tabular}{ |m{3.0cm}||c|c|c|c|c| }
 \hline
 NIST SARD Test Case & \scriptsize{Llama-2-70b-chat-hf} & \scriptsize{zephyr-7b-alpha} & \scriptsize{zephyr-7b-beta} & \scriptsize{Mistral-7B-Instruct-v0.1} & \scriptsize{Turdus} \\ 
 \hline
 \hline
 1792 (CWE-79) & 0 & 9 & 3 & 48 & 120 \\ 
 \hline
 500843 (CWE-476) & 181  & 8 & 10 & 9 & 60 \\ 
 \hline
 1779 (CWE-463) & 284 & 3 & 0 & 13 & 13 \\ 
 \hline
 1645 (CWE-20) & 1  & 23 & 6 & 156 & 157 \\ 
 \hline
 149165 (CWE-121) & 627  & 101 & 33 & 399 & 274 \\ 
 \hline
 2015 (CWE-329) &  328 & 12 & 5 & 94 & 46 \\ 
 \hline
 500757 (CWE-787) &  268 & 89 & 47 & 169 & 98 \\ 
 \hline
 149135 (CWE-489) &  19 & 6 & 1 & 0 & 17 \\ 
 \hline
 149203 (CWE-416) &  0 & 0 & 0 & 0 & 0 \\ 
 \hline
 149111 (CWE-134) &  59 & 224 & 48 & 36 & 281 \\ 
 \hline
 149185 (CWE-391) &  543 & 94 & 15 & 5 & 7 \\ 
 \hline
 1494 (CWE-74) &  950 & 188 & 47 & 155 & 368 \\ 
 \hline
 149179 (CWE-401) & 0  & 24 & 10 & 0 & 18 \\ 
 \hline
 149107 (CWE-415) & 0  & 6 & 2 & 2 & 0 \\ 
 \hline
 148871 (CWE-188) &  0 & 0 & 0 & 0 & 0 \\ 
 \hline
 149143 (CWE-120) &  686 & 144 & 42 & 26 & 621 \\ 
 \hline
 149103 (CWE-367) &  0 & 1 & 0 & 0 & 1 \\ 
 \hline
 1501 (CWE-822) & 3  & 5 & 0 & 0 & 3 \\ 
 \hline
 149199 (CWE-412) & 0  & 2 & 0 & 0 & 0 \\ 
 \hline
 149183 (CWE-468) &  158 & 162 & 90 & 19 & 151 \\ 
 \hline
 149085 (CWE-244) & 12  & 21 & 8 & 14 & 6 \\ 
 \hline
 2079 (CWE-464) &  3 & 5 & 2 & 0 & 0 \\ 
 \hline
 72 (CWE-248) & 86  & 3 & 6 & 2 & 16 \\ 
 \hline
 2046 (CWE-259) & 0  & 0 & 0 & 0 & 0 \\ 
 \hline
 2060 (CWE-457) &  0 & 3 & 1 & 0 & 1 \\ 
 \hline
 1989 (CWE-89) &  403 & 13 & 6 & 0 & 3 \\ 
 \hline
 114 (CWE-532) &  0 & 2 & 0 & 0 & 0 \\ 
 \hline
 20 (CWE-118) &  3 & 7 & 3 & 1 & 5 \\ 
 \hline
 1887 (CWE-22) & 0  & 0 & 0 & 0 & 0 \\ 
 \hline
 313 (CWE-119) & 941  & 258 & 59 & 54 & 416 \\ 
 \hline
 2016 (CWE-378) & 0  & 1 & 0 & 0 & 0 \\ 
 \hline
 149125 (CWE-122) & 2  & 59 & 3 & 7 & 67 \\ 
 \hline
 501035 (CWE-563) &  20 & 16 & 2 & 16 & 6 \\ 
 \hline
 1852 (CWE-170) & 3  & 24 & 8 & 1 & 42 \\ 
 \hline
 149163 (CWE-99) &  0 & 0 & 0 & 0 & 46 \\ 
 \hline
 149241 (CWE-78) & 0  & 4 & 0 & 0 & 11 \\ 
 \hline
\end{tabular}}
\caption{Count of unique and correct identification of the vulnerable line of source code, but not necessarily correctly identified CWE number (or being qualified as being present, or not present). The maximum possible quantity reported in each cell is $1000$. }
\label{table:code_line_identification_only}
\end{table*}

\begin{table*}[p!]
\centering
\begin{tabular}{ |c||m{8.8cm}| }
 \hline
 NIST SARD Test Case & Best performing GPT model and inference temperature (recall, precision, proportion of parseability) \\ 
 \hline
 1792 (CWE-79) & zephyr-7b-beta $0.4$: ($0.125$, $0.0667$, $0.08$)  \\ 
 \hline
 500843 (CWE-476) &  Turdus $0.1$: ($0.319$, $0.6$, $0.47$)  \\ 
 \hline
 1779 (CWE-463) & * \\ 
 \hline
 1645 (CWE-20) & zephyr-7b-beta 0.4: (0.611, 0.297, 0.18)  \\ 
 \hline
 149165 (CWE-121) & Llama-2-70b-chat-hf 0.1: (1.0, 1.0, 1.0)   \\ 
 \hline
 2015 (CWE-329) & zephyr-7b-beta 0.7: (0.077, 0.048, 0.13)  \\ 
 \hline
 500757 (CWE-787) & Turdus 0.9: (0.183, 0.177, 0.6)  \\ 
 \hline
 149135 (CWE-489) & * \\ 
 \hline
 149203 (CWE-416) & zephyr-7b-alpha 0.8: (0.24, 0.12, 0.25) \\ 
 \hline
 149111 (CWE-134) & Llama-2-70b-chat-hf 0.5: (0.727, 0.675, 0.77) \\ 
 \hline
 149185 (CWE-391) & * \\ 
 \hline
 1494 (CWE-74) & Mistral-7B-Instruct-v0.1 0.9: (0.029, 0.033, 0.67) \\ 
 \hline
 149179 (CWE-401) & zephyr-7b-beta 0.3: (0.25, 0.167, 0.16) \\ 
 \hline
 149107 (CWE-415) & zephyr-7b-alpha 0.2: (0.064, 0.029, 0.31) \\ 
 \hline
 148871 (CWE-188) & * \\ 
 \hline
 149143 (CWE-120) & Turdus 1.0: (0.237, 0.229, 0.8) \\ 
 \hline
 149103 (CWE-367) & zephyr-7b-alpha 0.6: (0.111, 0.083, 0.09) \\ 
 \hline
 1501 (CWE-822) & Mistral-7B-Instruct-v0.1 0.7: (0.014, 0.015, 0.73) \\ 
 \hline
 149199 (CWE-412)    & Mistral-7B-Instruct-v0.1 0.8: (0.019, 0.02, 0.53) \\ 
 \hline
 149183 (CWE-468)    & * \\ 
 \hline
 149085 (CWE-244)    & * \\ 
 \hline
 2079 (CWE-464)    & Turdus 0.9: (0.0159, 0.0175, 0.63) \\ 
 \hline
 72 (CWE-248)    & * \\ 
 \hline
 2046 (CWE-259)    & Llama-2-70b-chat-hf 0.1: (0.24, 0.0559, 1.0) \\ 
 \hline
 2060 (CWE-457)    & zephyr-7b-beta 0.5: (0.1, 0.0556, 0.1) \\ 
 \hline
 1989 (CWE-89)    & Llama-2-70b-chat-hf 0.1: (1.0, 1.0, 1.0) \\ 
 \hline
 114 (CWE-532)    & zephyr-7b-alpha 0.9: (0.0645, 0.0689, 0.31) \\ 
 \hline
 20 (CWE-118)    & * \\ 
 \hline
 1887 (CWE-22)    & Llama-2-70b-chat-hf 0.1: (1.0, 0.304, 1.0) \\ 
 \hline
 313 (CWE-119)    & Mistral-7B-Instruct-v0.1 0.1: (0.48, 0.48, 0.77) \\ 
 \hline
 2016 (CWE-378)    & * \\ 
 \hline
 149125 (CWE-122)    & zephyr-7b-beta 0.5: (0.46, 0.15, 0.13) \\ 
 \hline
 501035 (CWE-563)    & * \\ 
 \hline
 1852 (CWE-170)    & * \\ 
 \hline
 149163 (CWE-99)    & Mistral-7B-Instruct-v0.1 0.8: (0.0196, 0.026, 0.51) \\ 
 \hline
 149241 (CWE-78)    & Llama-2-70b-chat-hf 0.1: (1.0, 0.5, 1.0) \\ 
 \hline
\end{tabular}
\caption{Best Recall and Precision binary classification metrics (specifically the best mean of the recall and precision measures), in terms of GPT model and inference temperature, for each CWE NIST SARD test case. Data presentation is GPT model name: inference temperature (recall, precision, parsing rate). These metrics are taken from the dataset of $100$ inference calls (for the best performing GPT model and inference temperature). The proportion of the $100$ GPT outputs that were able to be parsed and to have a binary vector extracted is also reported as the third real number. The closer these three metrics are to $1.0$, the better the GPT model performed. The proportion of the outputs that could be parsed are not used for computing the best performing model and inference temperature, but it is useful to show the parse-ability success rate. These recall and precision metrics are quantifying the GPT models ability to correctly identify both the CWE number and the vulnerable line of source code. * denotes that all recall and precision metrics were $0$ for all GPT models and settings. Note that there is occasionally a complication that for all $100$ GPT output strings there are no parse-able outputs, but in these cases there are just no metrics that can be computed. If any of the settings resulted in the same F1 score, the GPT configuration with higher parsability rates is reported. }
\label{table:best_performing_recall_and_precision_measures}
\end{table*}

\begin{table*}[h!]
\centering
\scalebox{0.85}{
\begin{tabular}{ |m{3.0cm}||m{2.4cm}|m{2.3cm}|m{2.3cm}|m{2.4cm}|m{2.3cm}| }
 \hline
 NIST SARD Test Case & \scriptsize{Llama-2-70b-chat-hf} & \scriptsize{zephyr-7b-alpha} & \scriptsize{zephyr-7b-beta} & \scriptsize{Mistral-7B-Instruct-v0.1} & \scriptsize{Turdus} \\ 
 \hline
 \hline
 1792 (CWE-79) & CWE-134~452x & CWE-89~29x & CWE-20~32x & CWE-134~35x & CWE-476~164x \\ 
 \hline
 500843 (CWE-476) & CWE-123~333x & CWE-120~6x & CWE-20~77x & CWE-124~31x & CWE-126~46x \\ 
 \hline
 1779 (CWE-463) & CWE-123~373x  & CWE-121~24x & CWE-121~28x & CWE-124~64x &  CWE-126~135x \\ 
 \hline
 1645 (CWE-20) & CWE-78~395x  & CWE-78~221x  & CWE-78~51x & CWE-124~112x & CWE-78~437x \\ 
 \hline
 149165 (CWE-121) &  CWE-123~274x  & CWE-362~66x &  CWE-125~57x & CWE-116~143x &  CWE-126~197x \\ 
 \hline
 2015 (CWE-329) & CWE-327~357x  & CWE-382~45x & CWE-327~50x & CWE-125~47x & CWE-404~52x  \\ 
 \hline
 500757 (CWE-787) & CWE-121~474x  & CWE-476~39x & CWE-20~147x & CWE-124~106x & CWE-126~109x \\ 
 \hline
 149135 (CWE-489) &  CWE-123~281x & CWE-20~20x & CWE-20~49x & CWE-434~76x & CWE-89~198x \\ 
 \hline
 149203 (CWE-416) & CWE-125~327x  & CWE-125~85x & CWE-125~111x & CWE-129~130x & CWE-476~228x \\ 
 \hline
 149111 (CWE-134) & CWE-123~253x  & CWE-125~76x & CWE-125~42x & CWE-124~77x &  CWE-787~88x \\ 
 \hline
 149185 (CWE-391) & CWE-20~255x  & CWE-369~34x & CWE-20~66x & CWE-116~53x & CWE-369~201x \\ 
 \hline
 1494 (CWE-74) & CWE-123~860x  & CWE-121~100x & CWE-20~66x & CWE-124~137x & CWE-126~320x \\ 
 \hline
 149179 (CWE-401) &  CWE-121~403x & CWE-476~111x & CWE-121~41x & CWE-129~134x  & CWE-476~153x \\ 
 \hline
 149107 (CWE-415) &  CWE-123~245x & CWE-416~121x & CWE-416~24x & CWE-116~130x & CWE-405~149x \\ 
 \hline
 148871 (CWE-188) &  CWE-123~503x & CWE-434~27x & CWE-20~10x & CWE-134~84x & CWE-400~53x \\ 
 \hline
 149143 (CWE-120) &  CWE-123~696x & CWE-121~127x & CWE-121~74x & CWE-124~195x & CWE-126~372x \\ 
 \hline
 149103 (CWE-367) &  CWE-123~391x & CWE-125~21x & CWE-125~4x & CWE-134~69x & CWE-476~257x \\ 
 \hline
 1501 (CWE-822) & CWE-134~328x & CWE-125~22x & CWE-20~38x & CWE-129~119x & CWE-476~165x \\ 
 \hline
 149199 (CWE-412) &  CWE-23~677x & CWE-476~72x & CWE-20~54x & CWE-124~165x & CWE-476~100x \\ 
 \hline
 149183 (CWE-468) &  CWE-123~835x & CWE-125~74x & CWE-121~79x & CWE-116~137x & CWE-126~145x \\ 
 \hline
 149085 (CWE-244) & CWE-123~647x  & CWE-121~36x & CWE-121~93x & CWE-129~157x & CWE-120~209x \\ 
 \hline
 2079 (CWE-464) & CWE-123~501x  & CWE-125~79x & CWE-121~99x & CWE-124~156x & CWE-126~251x \\ 
 \hline
 72 (CWE-248) & CWE-123~357x  & CWE-125~12x CWE-121~12x & CWE-125~95x & CWE-124~42x & CWE-126~123x \\ 
 \hline
 2046 (CWE-259) & CWE-284~753x  & CWE-20~36x & CWE-20~102x & CWE-125~179x & CWE-89~255x \\ 
 \hline
 2060 (CWE-457) &  CWE-416~243x & CWE-416~29x & CWE-121~31x & CWE-124~86x & CWE-416~105x \\ 
 \hline
 1989 (CWE-89) & CWE-78~73x  & CWE-352~8x & CWE-20~31x & CWE-434~31x & CWE-476~46x \\ 
 \hline
 114 (CWE-532) & CWE-20~486x  & CWE-200~99x & CWE-20~107x & CWE-125~150x & CWE-20~119x \\ 
 \hline
 20 (CWE-118) & CWE-190~200x  & CWE-190~26x & CWE-190~39x & CWE-134~34x CWE-116~34x & CWE-190~85x \\ 
 \hline
 1887 (CWE-22) & CWE-20~758x  & CWE-125~44x & CWE-125~28x & CWE-129~57x & CWE-476~344x \\ 
 \hline
 313 (CWE-119) &  CWE-123~733x & CWE-121~130x & CWE-121~66x & CWE-124~147x & CWE-126~400x \\ 
 \hline
 2016 (CWE-378) &  CWE-22~140x & CWE-121~11x & CWE-20~73x & CWE-124~56x & CWE-404~119x \\ 
 \hline
 149125 (CWE-122) & CWE-123~615x  & CWE-125~57x & CWE-121~40x & CWE-124~106x & CWE-476~124x \\ 
 \hline
 501035 (CWE-563) &  CWE-20~593x & CWE-125~55x & CWE-20~47x & CWE-129~61x & CWE-476~73x \\ 
 \hline
 1852 (CWE-170) & CWE-123~802x & CWE-125~48x & CWE-125~68x & CWE-124~69x & CWE-476~203x \\ 
 \hline
 149163 (CWE-99) &  CWE-123~244x & CWE-434~71x & CWE-20~45x & CWE-124~67x & CWE-434~99x \\ 
 \hline
 149241 (CWE-78) & CWE-20~715x & CWE-120~25x & CWE-121~37x & CWE-124~11x & CWE-476~68x \\ 
 \hline
\end{tabular}}
\caption{Distribution of False Positives; the most frequent incorrectly identified CWE's (labeled as being \emph{Present} in the returned JSON datastructure) for each of the GPT models and each of the NIST SARD test cases. The frequency counts are from the totals out of the $10$ temperature settings, meaning that the maximum possible count in each cell is 1000x. The proportion of the GPT output that gave that incorrect CWE is reported to the right of the CWE number. The counts are obtained from all parse-able CWE entries, not just those that identify only a single CWE. If there are multiple CWEs with most frequency counts, all of the CWEs with the same most frequent counts are shown.  }
\label{table:most_common_incorrect_CWE_ids}
\end{table*}

\subsection{Example Correctly Identified Vulnerable Lines of Code}
\label{section:results_example_correct_JSON}

This section enumerates several parsed JSON outputs from the various GPT models and code test cases. These outputs were hand selected to be representative of cases that are correct (in the sense that the CWE number is correctly identified, as is the vulnerable line of source code), but whose description fields are in some cases correct and in other cases incorrect. Showing explicit examples of the output is intended to primarily show the variability of the generated \emph{Description} field, and also to show concretely what successful static code analysis identification looks like from the GPT models. 

No text changes were made to these GPT outputs except for formatting to fit within the text boxes. Note that these outputs are specifically examples of text from the enclosed triple backtick fields - any additional prose or output from the model is not shown.

\begin{tcolorbox}[title=Example of correct CWE number and code line identification in a JSON datastructure format from NIST SARD Test Case 500843. Produced from the Turdus GPT model with temperature 0.8. Note that the generated description is partially correct - CWE-476 is NULL pointer dereference but the description also mentions Divison-by-Zero which is not correct. ]
\begin{verbatim}
{'findings': [{'CWE_Number': 'CWE-476', 
'Status': 'Present', 
'Source_Code': 'first->str();', 
'Description': "Summary of the category of software vulnerability (Division By Zero) 
due to potential NULL pointer dereference when calling str() on 'first'"}]}
\end{verbatim}
\end{tcolorbox}

\begin{tcolorbox}[title=Example of correct CWE number and code line identification in a JSON datastructure format from NIST SARD Test Case 500843. Produced from the zephyr-7b-alpha GPT model with temperature 0.7. Note that the generated description is entirely incorrect.]
\begin{verbatim}
{'findings': [{'CWE_Number': 'CWE-476', 
'Status': 'Present', 
'Source_Code': 'first->str();', 
'Description': "The function 'f' takes a linked list of Tokens as an argument. 
If the last token in the list is a string, that string will be printed to stdout. 
This code is vulnerable to a format string attack because the 'f' function is using 
the'str' method of the last token in the list, which could contain a format string. 
An attacker can pass a format string as an argument to this function and potentially 
execute arbitrary code or cause a Denial of Service."}]}
\end{verbatim}
\end{tcolorbox}

\begin{tcolorbox}[title=Example of correct CWE number and code line identification in a JSON datastructure format from NIST SARD Test Case 500843. Produced from the Turdus GPT model with temperature 1.0. Note that the generated description is entirely incorrect. ]
\begin{verbatim}
{'findings': [{'CWE_Number': 'CWE-476', 
'Status': 'Present', 
'Source_Code': 'first->str();', 
'Description': "This code is vulnerable as it executes an untrusted input without proper 
validation before calling'str()'. This may result in memory corruption issues 
such as buffer overflow."}]}
\end{verbatim}
\end{tcolorbox}

\begin{tcolorbox}[title=Example of correct CWE number and code line identification in a JSON datastructure format from NIST SARD Test Case 500843. Produced from the Llama-2-70b-chat-hf GPT model with temperature 0.8. Note that the provided description is mostly correct in that the correct vulnerability name was identified but there is not a function named f in the source code. ]
\begin{verbatim}
{'findings': [{'CWE_Number': 'CWE-476', 
'Status': 'Present', 
'Source_Code': 'first->str();', 
'Description': 'Potential null pointer dereference vulnerability in function f'}]}
\end{verbatim}
\end{tcolorbox}

\begin{tcolorbox}[title=Example of correct CWE number and code line identification in a JSON datastructure format from NIST SARD Test Case 1792. Produced from the Turdus GPT model with temperature 0.3. Note that the generated description is correct.]
\begin{verbatim}
{'findings': 
[{'CWE_Number': 'CWE-79', 
'Status': 'Present', 
'Source_Code': 'fprintf(cgiOut, "Value number %d = %s<br />", i, q[i]);', 
'Description': 'This code is vulnerable to Cross-site Scripting (XSS) attacks due to 
insufficient input validation and sanitization.'}]}
\end{verbatim}
\end{tcolorbox}

\begin{tcolorbox}[title=Example of correct CWE number and code line identification in a JSON datastructure format from NIST SARD Test Case 1792. Produced from the Turdus GPT model with temperature 0.4. Note that the generated description is correct.]
\begin{verbatim}
{'findings': [{'CWE_Number': 'CWE-79', 
'Status': 'Present', 
'Source_Code': 'fprintf(cgiOut, "Value number %d = %s<br />", i, q[i]);', 
'Description': 'The code is vulnerable to Improper Neutralization of 
Input During Web Page Generation (CWE-79), as it uses sprintf() without 
validating or neutralizing user input in the web page generation process.'}]}
\end{verbatim}
\end{tcolorbox}

\begin{tcolorbox}[title=Example of correct CWE number and code line identification in a JSON datastructure format from NIST SARD Test Case 1792. Produced from the Turdus GPT model with temperature 0.4. Note that the generated description is correct except that the identified print statement in the description text is incorrect.]
\begin{verbatim}
{'findings': [{'CWE_Number': 'CWE-79', 
'Status': 'Present', 
'Source_Code': 'fprintf(cgiOut, "Value number %d = %s<br />", i, q[i]);', 
'Description': 'This code is vulnerable as it uses sprintf() on user-supplied 
input without proper validation or sanitization, which can lead to 
security risks such as Cross-Site Scripting (XSS)'}]}
\end{verbatim}
\end{tcolorbox}

\begin{tcolorbox}[title=Example of correct CWE number and code line identification in a JSON datastructure format from NIST SARD Test Case 1645. Produced from the Turdus GPT model with temperature 0.5. The generated Description field in this example is correct. ]
\begin{verbatim}
{'findings': [{'CWE_Number': 'CWE-20', 
'Status': 'Present', 
'Source_Code': 'system(buf);', 
'Description': "This code falls victim to Improper Input Validation where user input isn't 
validated before being used in a system() call, potentially allowing malicious 
commands to be executed."}]}
\end{verbatim}
\end{tcolorbox}

\begin{tcolorbox}[title=Example of correct CWE number and code line identification in a JSON datastructure format from NIST SARD Test Case 1645. Produced from the zephyr-7b-alpha GPT model with temperature 0.9. The generated Description field in this example is correct. ]
\begin{verbatim}
{'findings': [{'CWE_Number': 'CWE-20', 
'Status': 'Present', 
'Source_Code': 'system(buf);', 
'Description': 'The application uses system() to execute a user-supplied string without 
sanitizing the input, which is a potential injection vulnerability (CWE-20).'}]}
\end{verbatim}
\end{tcolorbox}

\begin{tcolorbox}[title=Example of correct CWE number and code line identification in a JSON datastructure format from NIST SARD Test Case 149165. Produced from the Llama-2-70b-chat-hf GPT model with temperature 0.9. The generated Description field in this example is entirely incorrect. ]
\begin{verbatim}
{'findings': [{'CWE_Number': 'CWE-121', 
'Status': 'Present', 
'Source_Code': "buffer[plop()] = '!';", '
Description': 'The use of a weak random number generator can lead to predictable output, 
which can be exploited by an attacker to execute arbitrary code.'}]}
\end{verbatim}
\end{tcolorbox}

\begin{tcolorbox}[title=Example of correct CWE number and code line identification in a JSON datastructure format from NIST SARD Test Case 149165. Produced from the Llama-2-70b-chat-hf GPT model with temperature 0.1. The generated Description field in this example is correct. ]
\begin{verbatim}
{'findings': [{'CWE_Number': 'CWE-121', 
'Status': 'Present', 
'Source_Code': "buffer[plop()] = '!';", 
'Description': "Buffer overflow vulnerability in the plop() function. 
The buffer 'buffer' is not large enough to hold the resulting value, 
which can cause a buffer overflow."}]}
\end{verbatim}
\end{tcolorbox}

\begin{tcolorbox}[title=Example of correct CWE number and code line identification in a JSON datastructure format from NIST SARD Test Case 149165. Produced from the Mistral-7B-Instruct-v0.1 GPT model with temperature 0.6. The generated Description field in this example is correct. ]
\begin{verbatim}
{'findings': [{'CWE_Number': 'CWE-121', 
'Status': 'Present', 
'Source_Code': "buffer[plop()] = '!';", 
'Description': 'Buffer overflow vulnerability'}]}
\end{verbatim}
\end{tcolorbox}

\begin{tcolorbox}[title=Example of correct CWE number and code line identification in a JSON datastructure format from NIST SARD Test Case 149165. Produced from the zephyr-7b-beta GPT model with temperature 0.2. The generated Description field in this example is correct. ]
\begin{verbatim}
{'findings': [{'CWE_Number': 'CWE-121', 
'Status': 'Present', 
'Source_Code': "buffer[plop()] = '!';", 
'Description': 'Buffer overflow vulnerability due to writing beyond the bounds 
of the buffer.'}]}
\end{verbatim}
\end{tcolorbox}

\begin{tcolorbox}[title=Example of correct CWE number and code line identification in a JSON datastructure format from NIST SARD Test Case 149165. Produced from the zephyr-7b-beta GPT model with temperature 0.4. The generated Description field in this example is not correct. ]
\begin{verbatim}
{'findings': [{'CWE_Number': 'CWE-121', 
'Status': 'Present', 
'Source_Code': "buffer[plop()] = '!';", 
'Description': 'Use of hard-coded values in security-sensitive contexts'}]}
\end{verbatim}
\end{tcolorbox}

\begin{tcolorbox}[title=Example of correct CWE number and code line identification in a JSON datastructure format from NIST SARD Test Case 500757. Produced from the Turdus GPT model with temperature 0.8. The generated Description field in this example is correct. ]
\begin{verbatim}
{'findings': [{'CWE_Number': 'CWE-787', 
'Status': 'Present', 
'Source_Code': 'p[10] = 7;', 
'Description': "This code falls under Out-of-Bounds Write category (CWE-787), 
as it attempts to write a value to an index (10) beyond the allocated 
memory space of 'p', potentially causing undefined behavior."}]}
\end{verbatim}
\end{tcolorbox}

\begin{tcolorbox}[title=Example of correct CWE number and code line identification in a JSON datastructure format from NIST SARD Test Case 1989. Produced from the Turdus GPT model with temperature 0.9. The generated Description field in this example is correct. ]
\begin{verbatim}
{'findings': [{'CWE_Number': 'CWE-89', 
'Status': 'Present', 
'Source_Code': 'queryStr += (" OR Value = \'" + *iter + "\'");', 
'Description': 'This code suffers from the SQL injection vulnerability since user inputs 
are being concatenated directly into an SQL query without any sanitation or 
parameterized queries, making it vulnerable to malicious input manipulation.'}]}
\end{verbatim}
\end{tcolorbox}

\begin{tcolorbox}[title=Example of correct CWE number and code line identification in a JSON datastructure format from NIST SARD Test Case 149111. Produced from the Llama-2-70b-chat-hf GPT model with temperature 0.3. The generated Description field in this example is correct. ]
\begin{verbatim}
{'findings': [{'CWE_Number': 'CWE-134', 
'Status': 'Present', 
'Source_Code': 'printf(container.fmt);', 
'Description': 'The use of user-supplied data in a format string 
vulnerable to exploitation.'}]}
\end{verbatim}
\end{tcolorbox}

\begin{tcolorbox}[title=Example of correct CWE number and code line identification in a JSON datastructure format from NIST SARD Test Case 149111. Produced from the Mistral-7B-Instruct-v0.1 GPT model with temperature 0.7. The generated Description field in this example is correct. ]
\begin{verbatim}
{'findings': [{'CWE_Number': 'CWE-134', 
'Status': 'Present', 
'Source_Code': 'printf(container.fmt);', 
'Description': 'Format string vulnerability can lead to buffer overflow 
or other security vulnerabilities.'}]}
\end{verbatim}
\end{tcolorbox}

\begin{tcolorbox}[title=Example of correct CWE number and code line identification in a JSON datastructure format from NIST SARD Test Case 149241. Produced from the zephyr-7b-alpha GPT model with temperature 0.2. The generated Description field in this example is correct. ]
\begin{verbatim}
{'findings': [{'CWE_Number': 'CWE-78', 
'Status': 'Present', 
'Source_Code': 'if (system(command) < 0)', 
'Description': 'Command Injection vulnerability, where an attacker can inject malicious
commands into the command parameter, which can be executed by the system() function.'}]}
\end{verbatim}
\end{tcolorbox}

\begin{tcolorbox}[title=Example of correct CWE number and code line identification in a JSON datastructure format from NIST SARD Test Case 313. Produced from the Llama-2-70b-chat-hf GPT model with temperature 0.6. The generated Description field in this example is correct. ]
\begin{verbatim}
{'findings': [{'CWE_Number': 'CWE-119', 
'Status': 'Present', 
'Source_Code': 'strncpy(buf, src, 11 % i);', 
'Description': 'Buffer overflow vulnerability in strncpy() function'}]}
\end{verbatim}
\end{tcolorbox}

%%%%%%%%%%%%%%%%%%%%%%%%%%%%%%%%%%%%%%%%%%%%%%%%%%%%%%%%%%%%%%
%%%%%%%%%%%%%%%%%%%%%%%%%%%%%%%%%%%%%%%%%%%%%%%%%%%%%%%%%%%%%%
%%%%%%%%%%%%%%%%%%%%%%%%%%%%%%%%%%%%%%%%%%%%%%%%%%%%%%%%%%%%%%
\section{Discussion and Conclusion}
\label{section:conclusion}

This study has examined the task of correctly identifying software weaknesses and vulnerabilities in software using open source Generative Pre-Trained Transformer (GPT) models. This was able to be examined because of two specific reasons - the first is the existence of the NIST SARD dataset, which is a very well labelled dataset that includes the vulnerable line of code and the relevant CWE. The second is the specific prompting template usage that allows fully automated extraction of JSON output for the vulnerability detection - and therefore error rates can be quantified for a large range of NIST SARD test cases, GPT models, and GPT model settings. Thus, this study reports a systematic analysis of automated vulnerability detection using GPT models. 

In general, our findings are that the correct CWE identification rate varies significantly depending on the specific code test case that is analyzed. For some test cases the correct CWE identification (and relevant line code of syntax) is surprisingly high, but for other test cases the identification accuracy rate is always 0 for all GPT models and temperature settings. In general, the error rates of this vulnerability detection is too high -- in particular, the false positive rate is very high. However, it is notable that for some cases, seemingly the more popular classes of CWEs that are mentioned in the type of internet text on which these GPT models were trained, the identification rates are quite good. As specific examples of the lowest error rate results (from Table \ref{table:best_performing_recall_and_precision_measures}), we list several specific examples of CWE identification binary classification accuracy rates (without considering rates for correctly identifying the vulnerable line of source code). Test case 1989 had a best recall of $0.928$, a best precision of $0.709$ (with a parsability rate of $0.42$) when analyzed with \texttt{Llama-2-70b-chat-hf} GPT model and temperature $0.1$. Test case 149165 had a best recall of $1.0$, a best precision of $1.0$ (with a parsability rate of $1.0$) when analyzed with \texttt{Llama-2-70b-chat-hf} GPT model and temperature $0.1$. Test case 149111 had a best recall of $0.727$, a best precision of $0.675$ (with a parsability rate of $0.77$) when analyzed with \texttt{Llama-2-70b-chat-hf} GPT model and temperature $0.5$. Test case 149241 had a best recall of $1.0$, a best precision of $0.5$ (with a parsability rate of $1.0$) when analyzed with \texttt{Llama-2-70b-chat-hf} GPT model and temperature $0.1$. Given the high accuracy rates of at least some of these test cases, it seems plausible that some of the current open source GPT models, if thorough verification is used, could correctly identify certain classes of vulnerabilities, such as buffer overflows, with very low error rates.

It should be emphasized that this particular software security analysis task requires a number of semantic connections to be made in the GPT model inference process (and of course, the correct specification of the JSON syntax) in order for the (automated) vulnerability identification to be correct. It is not clear to what extent a pre-trained language model can learn these semantic connections - but it is possible least for some of the CWEs, these connections were learned in the standard open source GPT models that we tested. This is is a non-trivial finding, and suggests that GPT models do have some capability to identify security vulnerabilities. However, it also needs to be emphasized that such usage of current GPT models should not be used in cases where the accuracy can not be tested. In other words, these GPT models and vulnerability identification prompting should not be used in software engineering tasks -- they should be used in cases where careful and precise testing of the GPT output can be performed for the purpose of examining specific characteristics of source code of interest. The primary reason for this is that the false positive rate is high - and not only is it high but the false positives are biased towards incorrectly identifying specific CWEs. In other words, simply measuring which CWE is predicted most frequently in a GPT output distribution (for a non-zero inference temperature) will typically not be correct.

The primary limitation with increasing the scope of more code test cases to more thoroughly test the capabilities of GPT models for software security analysis is the total amount of compute time that is required for performing these inference calls, especially in a systematic study of varying inference parameters. Future studies that increase the number of labeled source code examples (e.g., from the extensive amount of data available in the NIST SARD dataset) that are used for testing the accuracy GPT automated vulnerability identification would be very relevant.

An aspect of this study that is unclear is \emph{what} details of the source code are the GPT models using to infer relevant CWE's in the source code. In particular, what semantic details are used to parse whether a CWE is more likely or less likely. In some cases, it could be that context, such as what types of libraries are used or what types of strings are being constructed (such as SQL query strings), is sufficient to reduce the total number of possibly applicable CWEs so that random guessing among those feasible candidates accounts for some of the GPT model's performance. The metrics in this study do not determine what components of the source code are important; this would require quantifying the explainability of GPT models \cite{10.1145/3639372, 10.1145/3529755, turpin2023language, wu2024usable, singh2024rethinking} for this task. We leave this study open to future research, and emphasize that this type of detailed explainability analysis is absolutely a critical component of study before these GPT models are incorporated into any type of vulnerability detection in production software development and testing.

\section{Acknowledgments}
\label{section:acknowledgments}

Sandia National Laboratories is a multi-mission laboratory managed and operated by National Technology \& Engineering Solutions of Sandia, LLC (NTESS), a wholly owned subsidiary of Honeywell International Inc., for the U.S. Department of Energy’s National Nuclear Security Administration (DOE/NNSA) under contract DE-NA0003525. This written work is authored by an employee of NTESS. The employee, not NTESS, owns the right, title and interest in and to the written work and is responsible for its contents. Any subjective views or opinions that might be expressed in the written work do not necessarily represent the views of the U.S. Government. The publisher acknowledges that the U.S. Government retains a non-exclusive, paid-up, irrevocable, world-wide license to publish or reproduce the published form of this written work or allow others to do so, for U.S. Government purposes. The DOE will provide public access to results of federally sponsored research in accordance with the DOE Public Access Plan.

\clearpage

\setlength\bibitemsep{0pt}
\printbibliography

\appendix

\section{Markdown Computer Code Language Identifiers}
\label{section:appendix_markdown_code_identifiers}

\texttt{sql, c++, javascript, cpp, python, csharp, c, json, less, bash, sh, java, text, json... , json..., diff, css, assembly, xml, perl, yaml, css, scss, html, makefile, js, csv, lua, kotlin, arduino, javascript, csharp, rust, shellscript, erb, vbnet, `json, go, plaintext, php, instructions:, ocl, shell, json ?, `json`...}

\section{NIST SARD Test Cases}
\label{section:appendix_NIST_SARD_test_cases}

This section shows the exact source code for 10 of the NIST SARD test cases that are used in this study. The single vulnerable line of source code is highlighted as red text for each test case.

\noindent\begin{minipage}{.49\textwidth}
\begin{lstlisting}[caption={ NIST SARD Test case 500843, instance of CWE-476 },captionpos=b,label={SARD_source_code:500843},language=C,showstringspaces=false,style=base,keywordstyle=\color{blue}]

struct Token {
    const Token* nextArgument() const;
    const Token* next() const;
    int varId() const;
    void str() const;};
void f(const Token *first) {
    first = first->nextArgument();
    if (first)
        first = first->next();
    @first->str();@
}
\end{lstlisting}
\end{minipage}
\noindent\begin{minipage}{.49\textwidth}
\begin{lstlisting}[caption={ NIST SARD Test case 1792, instance of CWE-79. Note that some of the strings in this test case have been modified so that terms related to cross site scripting are removed. },captionpos=b,label={SARD_source_code:500843},language=C,showstringspaces=false,style=base,keywordstyle=\color{blue}]

#include <stdio.h>
#include <cgic.h>
#include <string.h>
#include <stdlib.h>

int cgiMain() 
{
	cgiHeaderContentType("text/html");
	 
	fprintf(cgiOut, "<html><head>\n");
	fprintf(cgiOut, "<title>Text: 1</title></head>\n");
	fprintf(cgiOut, "<body><h1>Text</h1>\n");
	 
	char q[4][1024];
	unsigned int i = 0;
	for (; i < 4; ++i){
		char name[2];
		sprintf(name,"q%d",i);
		cgiFormString(name, q[i], sizeof(q[i]));
		if (strlen(q[i]))
		{
			@fprintf(cgiOut, "Value number %d = %s<br />", i, q[i]);@
		}
	}		
	 
	fprintf(cgiOut, "</body></html>\n");
	return 0;
}

\end{lstlisting}
\end{minipage}

\noindent\begin{minipage}{.49\textwidth}
\begin{lstlisting}[caption={ NIST SARD Test case 1779, instance of CWE-463. },captionpos=b,label={SARD_source_code:1779},language=C,showstringspaces=false,style=base,keywordstyle=\color{blue}]

main ()
{
char *foo;
int counter;
foo=malloc(sizeof(char)*10);
for (counter=0;counter!=14;counter++){
@foo[counter]='a';@
printf("%s\n",foo);
}

\end{lstlisting}
\end{minipage}

\noindent\begin{minipage}{.49\textwidth}
\begin{lstlisting}[caption={NIST SARD Test case 1645, instance of CWE-20. },captionpos=b,label={SARD_source_code:1645},language=C,showstringspaces=false,style=base,keywordstyle=\color{blue}]

#include <stdio.h>
#include <stdlib.h>

#define	MAXSIZE		40

void test(char *str){
	char buf[MAXSIZE];
	snprintf(buf, sizeof buf, "/bin/echo %s", str);
	buf[MAXSIZE-1] = 0;
	@system(buf);@
}

int main(int argc, char **argv){
	char *userstr;
	if(argc > 1) {
		userstr = argv[1];
		test(userstr);
	}
	return 0;
}

\end{lstlisting}
\end{minipage}

\noindent\begin{minipage}{.49\textwidth}
\begin{lstlisting}[caption={NIST SARD Test case 149165, instance of CWE-121. },captionpos=b,label={SARD_source_code:149165},language=C,showstringspaces=false,style=base,keywordstyle=\color{blue}]

#include <stdlib.h>
#include <stdio.h>
#include <time.h>
#include <string.h>

unsigned int getRand()
{
	unsigned int r;
	FILE *f;
	f = fopen("/dev/urandom", "rb");
	if(f == NULL)
	{
		fprintf(stderr, "Error opening file\n");
		exit(-1);
	}
	if(fread(&r, sizeof r, 1, f) != 1)
	{
		fprintf(stderr, "Error reading file\n");
		fclose(f);
		exit(-1);
	}
	if(fclose(f) != 0)
		fprintf(stderr, "Error closing file\n");

	return r;
}

unsigned plop() {
	return getRand() % 256 + 127;
}

int main(int argc, char *argv[])
{
	char buffer[256];
	memset(buffer, 0, sizeof(buffer));
	@buffer[plop()] = '!';@		 
	printf("%s\n", buffer);
	return 0;
}

\end{lstlisting}
\end{minipage}

\noindent\begin{minipage}{.49\textwidth}
\begin{lstlisting}[caption={NIST SARD Test case 2015, instance of CWE-329. },captionpos=b,label={SARD_source_code:2015},language=C,showstringspaces=false,style=base,keywordstyle=\color{blue}]

#include <openssl/evp.h>

main(){
EVP_CIPHER_CTX ctx;
char key[EVP_MAX_KEY_LENGTH];
char iv[EVP_MAX_IV_LENGTH];
int b=8;
RAND_bytes(key, b);
memset(iv,0,EVP_MAX_IV_LENGTH);
@EVP_EncryptInit(&ctx,EVP_bf_cbc(), key,iv);@
return;
}

\end{lstlisting}
\end{minipage}

\noindent\begin{minipage}{.49\textwidth}
\begin{lstlisting}[caption={ NIST SARD Test case 500757, instance of CWE-787 },captionpos=b,label={SARD_source_code:500757},language=C,showstringspaces=false,style=base,keywordstyle=\color{blue}]
#include <cstdlib>
void f()
{
  char *p; p = (char *)malloc(10);
  @p[10] = 7;@
  free(p);
}

\end{lstlisting}
\end{minipage}

\noindent\begin{minipage}{.49\textwidth}
\begin{lstlisting}[caption={NIST SARD Test case 149135, instance of CWE-489. },captionpos=b,label={SARD_source_code:149135},language=C,showstringspaces=false,style=base,keywordstyle=\color{blue}]

#include <stdio.h>
#include <stdlib.h>
#include <stdbool.h>
#include <string.h>

static bool debug = false;

void promote_root() {
	if (debug) {
		 
		printf ("# You are root now...\n");
	}
}

int main(int argc, char *argv[])
{
	if (argc > 1)
	{
		unsigned i;
		const unsigned nbArgs = argc;

		for (i = 1; i < nbArgs; ++i)
		{@
			if (!strncmp(argv[i],"-debug",6)) @
			{
				debug = true;
				printf("Move to debugging mode\n");
				 
				if (strlen(argv[i]) >= 11 && !strncmp(argv[i]+6,":root",5))
				{
					promote_root();
				}
			}
		}	
	}
	else
	{
		printf("No args...\n");
	}

	return 0;
}


\end{lstlisting}
\end{minipage}

\noindent\begin{minipage}{.49\textwidth}
\begin{lstlisting}[caption={NIST SARD Test case 149203, instance of CWE-416. },captionpos=b,label={SARD_source_code:149203},language=C,showstringspaces=false,style=base,keywordstyle=\color{blue}]

#include <stdlib.h>
#include <stdio.h>
#include <string.h>

typedef struct str_t str;
struct str_t {
	union {
		int   a;
		char *b;
	} foo;
};

int main()
{
	str container;
	char *str1;
	container.foo.a = 0;
	
	if ((container.foo.b = (char *)malloc(256*sizeof(char))) != NULL)
	{
		strcpy(container.foo.b, "Falut!");
		container.foo.b[0] = 'S';
		printf("%s\n", container.foo.b);
		free(container.foo.b);
		str1 = malloc(65536); 
 		if (str1 == NULL)
        		{printf("Memory allocation problem"); return 1;}

		free(str1);
		
		if (container.foo.b)
			@{container.foo.b[0] = 'S'; printf("%s\n", container.foo.b);}@	 
	}
	return 0;
}

\end{lstlisting}
\end{minipage}

\noindent\begin{minipage}{.49\textwidth}
\begin{lstlisting}[caption={NIST SARD Test case 149111, instance of CWE-134. },captionpos=b,label={SARD_source_code:149111},language=C,showstringspaces=false,style=base,keywordstyle=\color{blue}]

#include <stdlib.h>
#include <stdio.h>
#include <string.h>

#define MAX_SIZE 255

typedef struct cont_t cont;
struct cont_t {
	char fmt[MAX_SIZE];
};

int main(int argc, char *argv[])
{
	if (argc > 1) {
		cont container = {.fmt=""};
		strncpy(container.fmt, argv[1],MAX_SIZE-1);
		container.fmt[MAX_SIZE - 1] = '\0';
		@printf(container.fmt);		@		 
	}
	return 0;
}

\end{lstlisting}
\end{minipage}

\end{document}